\documentstyle[aps,twocolumn,prb,epsf]{revtex} %with lines(3) of '%+++++'

\newcommand{\av}[1]{\left< #1 \right>}

\newcommand{\K}{\mathrm{\ K}}
\newcommand{\V}{\mathrm{\ V}}
\newcommand{\Vs}{\tilde V}

\renewcommand{\vec}[1]{\mathbf{#1}}

\begin{document}
\draft
\twocolumn[    
          %+++++
\hsize\textwidth\columnwidth\hsize\csname @twocolumnfalse\endcsname    %+++++

\title{Thermoelectric power of Cerium and Ytterbium intermetallics}

\author{V. Zlati\'c$^1$, 
I. Milat$^{1,2}$, 
B. Coqblin$^3$ 
and G. Czycholl$^4$\\[2ex]
$^1$Institute of Physics, Bijeni\v cka cesta 46, 
P. O. Box 304, HR-10001 Zagreb, Croatia\\         
$^2$Institute for Theoretical Physics, ETH-H\"ongerberg, 
CH-8093 Z\"urich, Switzerland\\
$^3$Universite Paris-Sud, Bât. 510, Centre d'Orsay,
91405 - Orsay - Cedex, France\\         
$^4$Institute for Theoretical Physics, Bremen University, 
D-28334 Bremen, Germany\\
}

\date{\today}

\maketitle
\widetext

\begin{abstract}
Temperature dependence of the thermoelectric power (TEP) of metallic 
systems with cerium and ytterbium ions exhibits some characteristic features, 
which can be used to classify these systems into distinct categories. 
We explain the observed properties in terms of the Kondo effect 
modified by the crystal field (CF) splitting and relate various shapes 
to different energy scales that characterize scattering of conduction 
electrons on Ce and Yb ions at different temperatures. The low- and 
high-temperature behaviors are calculated for different fixed point models  
and the overall shape of the TEP is obtained by interpolation.
At high temperatures we use the Coqblin-Schrieffer model (CSM)  
and calculate the thermoelectric power by the perturbation 
expansion with renormalized coupling constants. 
Performing the renormalization by the 'poor man's scaling' (PMS), 
we find for large CF splitting two Kondo scales, $T_K$ and $T_K^H\gg T_K$. 
The thermopower obtained in such a way exhibits a large maximum 
at about $T_K^H$, and a sign change at about $T_K$. 
The PMS also shows that for $T\ll T_K^H$ the {\it f} level behaves 
as an effective multiplet with the degeneracy of the CF ground state and 
Kondo scale $T_K$. Thus, we assume that the low-temperature fixed 
point models do not require the CF splitting. 
The properties of dilute Ce and Yb alloys are obtained for $T\leq T_K$ 
from an effective spin-degenerate single-impurity Anderson model (SIAM). 
The parameters of SIAM are such that its effective Kondo scale 
coincides with the Kondo scale $T_K$ of the CSM. 
The stoichiometric compounds are described by an effective spin-degenerate 
periodic Anderson model (PAM), such that its characteristic energy 
scale $T_0$ is the same as $T_K$. 
The transport coefficients of PAM and SIAM are calculated by 
perturbation theory. 
The interpolation between the Anderson model results, valid for $T\leq T_K$, 
and the CSM, results valid for $T\geq T_K$, provides a qualitative 
explanation of the TEP of most Ce and Yb intermetallics. 
\end{abstract}

  % (thermoelectric and thermomagnetic effects)
 ]         % +++++

\narrowtext

%============================================================================
\subsection*{Introduction}
The thermo-electric properties of intermetallic compounds with Ce and Yb ions  
with one {\it f} electron or {\it f} hole show many interesting features and their 
thermoelectric power (TEP) considered as a function of temperature, $S(T)$, 
assumes a variety of shapes. 
\cite{steglich.77,petersen.79,levin.81,schneider.83,aliev.84,sparn.85,gottwick.85,amato.85,jaccard.85,jaccard.87,jaccard.87b,onuki.87a,sakurai.87,amato.88b,amato.88a,jaccard.88,gottwick.87,sakurai.88,sampathkumaran.89a,lees.90,casanova.90,jaccard.90,bauer.91,cibin.92,jaccard.92,bando.93,sakurai.95,jaccard.96,ocko.99,huo.00,sakurai.00,ocko.01a,ocko.01b} 
The function $S(T)$ is often non-monotonous and in some systems 
it exhibits a sign change at lowest temperatures.
Above 100 K, the TEP can assume giant values, and 
much of the recent interest in heavy fermion thermo-electricity 
is due to the belief that some of the new systems, 
with the thermopower larger than 150 $\mu$V/K, 
might be useful for application.\cite{mahan}
The functional form of $S(T)$ appears to be quite complicated,  
but systems with similar thermopower exhibit similarities 
in other thermodynamic and transport properties as well.
The shape of $S(T)$ can be used to classify Ce
and Yb intermetallics and alloys into distinct groups.\cite{jaccard.96}

The thermo-electric anomalies due to the rare earth ions have stimulated 
a lot of theoretical work.
\cite{peschel.70,coqblin.76,maekava.86,bickers.87,fisher.89,monnier.90,evans.92,zlatic.93,costi.94a,zlatic.82,zlatic.74,mahan.97}
It is now clear that most of the experimental results can 
be understood in terms of the exchange scattering of conduction 
electrons on the 4f$^1$ state of Ce or the 4f$^{13}$ state of Yb  
provided one takes into account the splitting of 
the {\it f} states due to the crystalline electric field (CF).
The complete description would have to consider the many-body effects 
due to the local correlation and the hybridization of {\it f} electrons 
with the conduction states, together with the point group symmetry of the 
crystal and the spin-orbit and the CF splittings. 
Although the solution of such a general model is not available, 
one can assume, using the usual renormalization group arguments, that 
the low- and high-temperature behaviors belong to different fixed points. 
Provided we know the solution of various models pertaining to 
different fixed points we can then construct the overall temperature 
dependence of $S(T)$ by interpolation. 
Here we show that the typical shapes of the thermopower found in metallic
systems with Ce and Yb ions can be explained in such a simple way.
The magnetic response of these systems has been explained recently 
using the same approach. 

At high temperatures we use the Coqblin-Schrieffer model (CSM), 
which takes into account the degeneracy and the CF splitting
of Ce 4f$^1$ electron and Yb 4f$^{13}$ hole. 
Thus, we describe the {\it f} ions as local magnetic moments and assume 
that the scattering of conduction electrons on such moments is 
incoherent regardless of their concentration. 
This is consistent with the observation that the high-temperature 
thermopowers of dilute alloys and stoichiometric compounds are often 
surprisingly similar.\cite{jaccard.87} 
The giant values of $S(T)$ around the high-temperature 
maximum are explained very well by the 3rd-order perturbation 
theory for the transport relaxation time.
\cite{coqblin.76,evans.92,coqblin.69,coqblin.72}  
The perturbation theory for the CSM can be further improved by 
renormalizing the coupling constants by the poor man's scaling.
\cite{anderson.70,nozieres.80}
The scaling solution\cite{yamada.84,hanzawa.85,hewson.93} shows
that a large CF splitting $\Delta$ leads to 
two Kondo temperatures, $T_K$ and $T_K^H\gg T_K$. 
The high-temperature maximum of $S(T)$ relates to $T_K^H$ and 
the reduction of temperature can lead to a sign change at $T_K$. 
Typical values of $T_K$, $T_K^H$, and $\Delta$ in Ce 
and Yb system are 10 K, 100 K, and 100 - 1000 K, respectively. 
However, the renormalized perturbation expansion becomes unreliable 
around $T_K$ and cannot be used to describe the non-monotonic 
thermopower found in many heavy fermions below $T_K$. 

To obtain the fixed point models appropriate for the description 
of the system at low-temperatures, we use the scaling result 
that the CF split {\it f} state of the CSM can be approximated for 
$T\ll T_K^H$ by an effective multiplet with the degeneracy of the 
CF ground state. 
Thus, as regards the low-energy dynamics, we assume that the sole 
effect of the excited CF states is to provide the right Kondo scale 
for an effective model. Once the lowest energy scale is set to $T_K$, 
the excited {\it f} states can be removed from the low-temperature problem. 
In addition, we assume that for $T\simeq T_K$ the scattering of conduction 
electrons is incoherent, regardless of the concentration of the rare 
earth ions. 
The properties of the system for $T\leq T_K$ are then found by 
considering the 4f ions as an assembly of {\it f} levels with the degeneracy 
of the CF ground state and no CF splitting. The transport and the 
thermodynamic anomalies are related to the scattering of conduction 
electrons on such effective {\it f} states. However, to allow for the 
possibility that for large concentration of Ce and Yb ions the 
ground state of the system could become coherent, we treat the 
dilute alloys and the stoichiometric compounds in a different way.

In dilute alloys, where the scattering of conduction electrons on 
rare earth ions remains incoherent down to lowest temperatures, 
the effective low-temperature problem for the Ce and Yb impurities 
is described by the single-impurity Anderson model (SIAM). 
The effective f level of the SIAM is assumed to have the 
same degeneracy as the CF ground state of the $4f^1$ level 
and the effective parameters are adjusted in such a way that 
the width of the {it f}-electron Kondo resonance\cite{hewson.93} 
coincides with the Kondo temperature $T_K$ of the CSM. 
Thus, we assume that for $T\leq T_K$ the CSM can be represented 
by the SIAM. The TEP of the SIAM is then found by some 
accurate method.\cite{costi.94a,zlatic.82}

In stoichiometric compounds, the problem is somewhat more difficult, 
because the reduction of temperature leads to the crossover to the 
coherent regime. We assume that for $T\leq T_K$ the properties of 
the lattice of Ce or Yb ions are described by a periodic Anderson model 
(PAM), such that the effective {\it f} levels have the same degeneracy 
as the lowest CF state of the 4f$^1$ or 4f$^{13}$ electrons. 
The absence of the excited CF states simplifies the 
calculations and the low-temperature transport coefficients 
of the PAM can be obtained by the self-consistent perturbation theory 
with respect to the correlation $U$.\cite{czycholl.91} 
This leads to a characteristic energy scale $T_0$, such that 
for $T\leq T_0$ the model describes a coherent Fermi-liquid state,  
and for $T\geq T_0$  an incoherent state. That is, above $T_0$, 
we consider the rare earth ions as independent scattering centers 
for conduction electrons. 
The effective parameters are now adjusted to give $T_0\simeq T_K$, 
and we assume that for $T\leq T_K$ the PAM and the CSM are 
dynamically equivalent. Above $T_K$, the physical state begins 
to deviate from the incoherent high-temperature state of the PAM 
which neglects the excited CF states. 
For $T\gg T_K$ the low-energy properties of stoichiometric 
compounds are described by the CSM with the CF splitting. 
Once we obtain the solution of various fixed point model, 
the overal temperature dependence is found by interpolation. 

The paper is organized as follows. First we describe the 
experimental results. Then we present the scaling solution 
of the CSM and discuss the TEP obtained by the renormalized 
perturbation theory. Next, we summarize the low-temperature results 
obtained for the SIAM and the PAM. Finally, we obtain the full TEP 
by interpolating between various fixed point solutions  
and use these results to discuss the data. 

\subsection*{Classification of the experimental data}

The experimental results for TEP of the Ce- and Yb-based intermetallic 
compounds exhibits some characteristic features, which can be used to divide 
these systems into several distinct groups.\cite{jaccard.96}

%%%%   a shape %%%%%%%%%
The type (a) thermopower is characterized by a deep negative minimum 
at low temperatures and a broad positive high-temperature peak between 100 K 
and 200 K. The type (a) behavior is found in a large majority of 
Ce-based heavy fermion compounds like 
CeCu$_2$Si$_2$,\cite{schneider.83,sparn.85,jaccard.85}
CePb$_3$\cite{gottwick.87},
CeCu$_2$Ge$_2$,\cite{gottwick.87,jaccard.92,jaccard.96} 
CePd$_2$Si$_2$,\cite{amato.85,jaccard.96}
CePdSn,\cite{huo.00}
CePdGe,\cite{sakurai.00} 
CeAl$_3$,\cite{sparn.85,jaccard.87} 
CeRh$_{2-z}$Ni$_z$Si$_2$ for small $z$,\cite{sampathkumaran.89a}
and CePtGe$_2$.\cite{sakurai.00}
It is also found in Yb-based systems like 
YbAgCu$_4$ 
or 
YbPd$_2$Si$_2$.\cite{casanova.90}

%%%%  shape b  %%%%%%%%%
The type (b) thermopower is similar to the type (a), except that
at very low temperatures the thermopower changes sign again and 
exhibits an additional positive peak. 
These features are often found 
in dilute Ce-alloys, like 
Ce$_x$La$_{1-x}$Ni,\cite{sakurai.87} 
Ce$_x$La$_{1-x}$Al$_3$,\cite{cibin.92} 
Ce$_{x}$La$_{1-x}$Pd$_2$Si$_2$,\cite{bando.93} 
and in some concentrated systems like 
Ce$_x$La$_{1-x}$\cite{petersen.79}
and  
Ce$_x$La$_{1-x}$Cu$_2$S$i_2$,\cite{aliev.84}
or stoichiometric compounds 
CeAl$_3$,\cite{sparn.85,jaccard.87} 
or YbAgCu$_4$.\cite{casanova.90} 
The additional low-temperature peak emerges in 
CePd$_2$Si$_2$ 
with pressure,\cite{jaccard.96} and in 
Ce$_{x}$La$_{1-x}$Pd$_2$Si$_2$ 
with chemical pressure.\cite{bando.93} 
In some cases the positive low-temperature peak 
is concealed by the superconducting or magnetic transitions. 
For example, in 
CeCu$_2$Si$_2$\cite{sparn.85} 
the positive low-temperature upturn does not appear 
in the zero-field data and it is only seen in an 
external magnetic field which suppresses the
superconducting transition. 

%%%%  shape c  %%%%%%%%%
In type (c) systems the low-temperature maximum is more pronounced 
than in type (b) but the separation to the high-temperature maximum 
is small and such that the minimum in-between does not extend down to  
negative values. The sign of $S(T)$ remains in type (c) systems 
the same at all temperatures. 
These features are found in dilute systems like 
Ce$_{x}$(La$_{1-z}$Y$_z$)$_{1-x}$Al$_2$,\cite{steglich.77}
Ce$_x$La$_{1-x}$Cu$_6$,\cite{onuki.87a} and 
Ce$_x$La$_{1-x}$Ru$_2$Si$_2$,\cite{amato.88b} 
in Ce-rich compounds like 
Ce(Pb$_{1-z}$Sn$_z$),\cite{sakurai.88} 
CeRu$_2$Si$_2$,\cite{amato.88a} 
Ce(Cu$_z$Au$_{1-z}$)$_6$ for small z,\cite{lees.90}
and also in 
Ce$_x$La$_{1-x}$Ni$_{0.8}$Pt$_{0.2}$,\cite{sakurai.95}  
Ce$_x$La$_{1-x}$Cu$_2$Si$_2$.\cite{ocko.99,ocko.01a} 
Ce$_x$Y$_{1-x}$Cu$_2$Si$_2$.\cite{ocko.01b} 

%%%%  shape d  %%%%%%%%%
The type (d) thermopower exhibits a large high-temperature peak, 
and perhaps a shoulder on the low-temperature side of that peak. 
This behavior is found at ambient pressure in 
CeCu$_6$ and Ce$_x$La$_{1-x}$Cu$_6$,\cite{onuki.87a} 
CeInCu$_2$,\cite{jaccard.90} 
CeCu$_3$Ga$_2$, and CeCu$_3$Al$_2$,\cite{bauer.91}
CePNiGe,\cite{sakurai.00}
and under high pressure in 
CeCu$_2$Si$_2$,\cite{jaccard.85} 
CeCu$_2$Ge$_2$,\cite{jaccard.96} and 
CePd$_2$Si$_2$.\cite{jaccard.96}

%%%%%%%%%%  e  %%%%%%%%
Finally, the TEP of shape (e) is found in {\em valence fluctuators}, 
where one observes a monotonic increase all the way up to room temperatures.
This is seen in 
CeBe$_{13}$,\cite{jaccard.85}
CePd$_3$,\cite{jaccard.87b} 
CeNi$_2Si_2$,\cite{sampathkumaran.89a,levin.81}
or in 
CeNi and CeNi$_2$.\cite{gottwick.85} 

The border between various categories is not very sharp and the TEP can be
transformed from one type to another by changing the environment of 4f ions.
In addition to the already listed examples, we mention 
Ce$_x$La$_{1-x}$Cu$_6$,\cite{onuki.87a} 
where for $x<0.5$ one finds two peaks 
separated by a well resolved minimum, while for $x>0.9$ only a single hump 
with a shoulder on the low-temperature side\cite{sakurai.95} remains. 
Similarly, by changing the concentration of Ce ions, one transforms $S(T)$ of 
Ce$_{x}$La$_{1-x}$Pd$_2$Si$_2$ \cite{bando.93} 
from type (a) to type (b), that of 
Ce$_{x}$La$_{1-x}$Cu$_2$Si$_2$\cite{aliev.84,ocko.99,ocko.99,ocko.01a} 
from type (a) to type (c), and that of 
Ce$_{x}$Y$_{1-x}$Cu$_2$Si$_2$\cite{ocko.01b} 
from type (a) to type (d).
Direct application of pressure (rather than chemical pressure) 
transforms the TEP of 
CePd$_2$Si$_2$,\cite{jaccard.96} 
CeCu$_2$Si$_2$ \cite{jaccard.85} 
or 
Ce Al$_3$,\cite{jaccard.88}
from type (a) or (b) to type (c) or (d).

Typical behaviors of the TEP are found in the 
Ce$_{1-x}$La$_x$Cu$_2$Si$_2$ family 
of intermetallic compounds for $0.01\ll x\ll 1.0$ at \% of Ce,\cite{ocko.99} 
where the TEP data exhibit nearly all the shapes discussed above. 
(The positive low-temperature peak which is seen in the dilute limit  
is also found for large concentration of Ce ions, provided one 
suppresses the superconducting transition by the magnetic field.)
The peaks in the thermoelectric power correlate with the 
logarithmic behavior in the electrical resistance;  
the magnetic susceptibility data\cite{aviani.01} indicate that above 100 K 
the local moment is 6-fold degenerate, while below 20 K it appears to 
be only 2-fold degenerate.  

In summary, the experimental data show that 
the characteristic shapes (a) -- (e) appear regardless 
of the concentration of magnetic ions. The high-temperature thermopower 
is insensitive to small changes in the concentration of rare earth ions 
and the thermoelectric anomalies are accompanied by the 
anomalies in other transport and thermodynamic properties. 
We take this behavior as an indication that the TEP anomalies 
are due to the scattering of conduction electrons from the 4-f state 
of Ce or Yb ions, and that the high-temperature scattering 
is incoherent regardless of the concentration of magnetic ions. 

%%%%%%%%%%%%%%%%%%%%%%%%%%%%%%%%%%%%%%%%%%%%%%%%%
\subsection*{High-temperature approximations                  
\label{model} }
%%%%%%%%%%%%%%%%%%%%%%%%%%%%%%%%%%%%%%%%%%%%%%%%%
The high-temperature properties of the rare earth ions in metallic 
environment are modeled Coqblin-Schrieffer Hamiltonian\cite{coqblin.69}  
with the CF splitting,  
\begin{eqnarray}
                                                    \label{CS_model}
  H =&&    \sum_\nu E_\nu a^{\dag}_\nu a_\nu
         + \sum_{k} \sum_\nu \epsilon_k\, c^{\dag}_{k\nu} c_{k\nu}
                                                          \nonumber\\
    &-& J_0 \sum_{kk'} \sum_{\nu\nu'} c^{\dag}_{k'\nu'} c_{k\nu}
             ( a^{\dag}_\nu a_{\nu'} 
              - \delta_{\nu\nu'} \langle{n_\nu}\rangle )  \nonumber\\
    &+& \sum_{k,k'} \sum_\nu (V_0 - J_0\langle{n_\nu}\rangle) 
           c^{\dag}_{k'\nu} c_{k\nu} 
\end{eqnarray}
where all the symbols have their usual meaning.
The first term describes the CF-split 4f$^1$ state of Ce ions 
or 4f$^{13}$ state of Yb ions, 
the second term describes the conduction band of width $2D_0$ and 
a constant density of states $\rho_0 $, the third term defines 
the pure (non-diagonal) exchange scattering between 4f$^1$ states and 
band electrons, and the last term is the total (diagonal) potential 
scattering. We consider here the antiferromagnetic coupling only.
The summation over $\nu$ is over all the CF states, and   
$J_0$ and $V_0$ are the initial coupling constants,
which are assumed to be $\nu$-independent.
For simplicity, we represent the 4f$^1$ electrons of Ce 
(the 4f$^{13}$ hole of Yb) by their lowest $J=5/2$ ($J=7/2$) 
spin-orbit state, 
and consider the CF scheme in which the ground state level at $E_m$ 
and the excited state at $E_M$ are $\alpha_m$-fold and $\alpha_M$-fold 
degenerate, respectively. The energy separation is $E_M-E_m=\Delta$. 
The model neglects the coherent scattering on Ce ions   
and is most appropriate for dilute alloys. However, it also applies 
to concentrated systems at temperatures such that the scattering of 
conduction electrons on the rare earth ions is incoherent.

%%%%%%%%%%%%%%%%%%%%%%%%%%%%%%%%%%%%%%%%%
\subsubsection*{Perturbation theory for the thermoelectric power
\label{transport}}
%%%%%%%%%%%%%%%%%%%%%%%%%%%%%%%%%%%%%%%%
We describe the high-temperature  
heat and charge transport by the Boltzmann equation  
and evaluate the scattering rate for the transport relaxation 
time up to the 3rd order in $J_0$.\cite{coqblin.72,coqblin.76,evans.92}
All the computational details can be found in the papers by 
Cornut and Coqblin\cite{coqblin.72} 
and Bhattacharjee and Coqblin,\cite{coqblin.76}
which we refer to as CC and BC, respectively. 
Here we just quote the thermopower result obtained by BC for 
the doublet-quartet CF scheme ($\alpha_m=2$, $\alpha_M=4$), 

\begin{equation}
                                  \label{eq:Sbc}
  S = \frac{k_B}{e} \rho_0 \frac{S_\Delta}{R_\Delta} 
  G_1(\Delta,0) , 
\end{equation}
where ${S_\Delta}$ and ${R_\Delta}$ are given by 
\begin{eqnarray}
%  (\ref{eq:C12}) +  (\ref{eq:B12})
 && S_\Delta
  = 
    16 |J_{Mm}|^2
    \left[
      \left(
        2\av{n_m}J_{m} + 4\av{n_M}J_{M}
      \right)
      \tanh\left(\frac{\Delta}{2T}\right)
    \right.
    \nonumber \\
    &&-
    \left.
      \left(
        \Vs_m + J_{m}\av{n_m} + \Vs_M + J_{M}\av{n_M}
      \right)
      \left(
        \av{n_m} - \av{n_M}
      \right)
    \frac{}{}
    \right] , 
                                  \label{eq:S_Delta}
\end{eqnarray}
\begin{eqnarray}
                                  \label{eq:R_Delta}
  R_\Delta &=& 
  2\left[
    \Vs_m^2 + 2 J_{m}^2 \av{n_m} \left(
                            1 - \frac{\av{n_m}}{2} 
                          \right)
  \right] 
  \nonumber\\
 &&+ 4\left[
    \Vs_M^2 + 4 J_{M}^2 \av{n_M} \left(
                            1 - \frac{\av{n_M}}{4}      
                          \right) 
  \right]
  \nonumber\\
  &&+ 16 |J_{Mm}|^2
       \left(
         \frac{\av{n_m}}{1+e^{\Delta/T}}
         +
         \frac{\av{n_M}}{1+e^{-\Delta/T}}
       \right), 
\end{eqnarray}
and 
\begin{equation}
  \label{eq:B.G1}
  G_1(\Delta, 0) 
  =
  \frac{\Delta}{T}
  \left[1 + \frac { \Delta}{2\pi T}\,
    Im\; \psi'\left(i \frac {\Delta}{2\pi T} \right) \right].
\end{equation}
The derivative of the psi-(digamma-) function\cite{abramowitz} 
is denoted by $\psi'(x)$, the occupancy of the CF states is 
\begin{equation}
  \label{eq:nmnM}
  \av{n_m} = \frac 1{ 2 + 4 e^{-\Delta/T}}, \;\;\;\;\;\;
  \av{n_M} = \frac {e^{-\Delta/T}}{ 2 + 4 e^{-\Delta/T}}, 
\end{equation}
and the effective potential scattering is 
\begin{equation}
                                \label{couplings}
   \Vs_m = V_m - J_m \av{n_m}, \;\;\;\;\
   \Vs_M = V_M - J_M \av{n_M}. 
\end{equation}
Note, $\rho_0{S_\Delta}/{R_\Delta}$ is a dimensionless quantity, 
$\rho_0$ is the density of conduction states arround the chemical 
potential $\mu$, and the prefactor in Eq.~(\ref{eq:Sbc}) is 
$k_B/e\approx 86 \mu V/K$).  

The qualitative features of the thermopower described by Eq.~(\ref{eq:Sbc}) 
with constant and isotropic coupling constants 
($J_m = J_M = J_{mM} \equiv J_0$, and $\Vs_m = \Vs_M \equiv \Vs$)
follow straightforwardly from the asymptotic expansion of 
the elementary  functions in Eq.~\ (\ref{eq:Sbc})  
and can be summarized as follows (detailed analysis 
can be found in Ref. BC).
At temperatures such that $T > \Delta$ the thermopower is small and 
behaves as $1/T$. 
The sign of $S$ and its slope depend on the relative size of $J_0$ and 
$\tilde V$. As temperature decreases, the thermopower increases up 
to a large (positive or negative) value at about $T_{max}\simeq \Delta/3$.
At temperatures such that $T \ll \Delta$ we find  $G_1 \propto T/\Delta$ 
and $S(T)\propto  T$. 

For parameters in the physical range one easily finds the peak value
of thermopower above 50 $\mu$V/ K and Eq.~(\ref{eq:Sbc}) captures the
essential high-temperature features of Ce and Yb
intermetallics.\cite{coqblin.76} 
At low temperatures, however, perturbative approach leading to 
Eq. \ (\ref{eq:Sbc}) fails and the observed, non-monotonic 
low-temperature behavior is not reproduced.

%%%%%%%%%%%%%%%%%%%%%%%%%%%%%%%%%%%%%%%%%
\subsubsection*{Poor man's scaling
\label{scaling}}
%%%%%%%%%%%%%%%%%%%%%%%%%%%%%%%%%%%%%%%%
The change of sign of the TEP below $T_{max}$, which is often 
seen in the experimental data, cannot be obtained from 
Eq.~(\ref{eq:Sbc}) for any value of $J_0$ and $V_0$. 
However, these features follow from the perturbation expansion, 
provided we renormalize the coupling constants by the poor man's scaling. 

\begin{figure}[tb]
  \epsfxsize=3.0in \epsffile{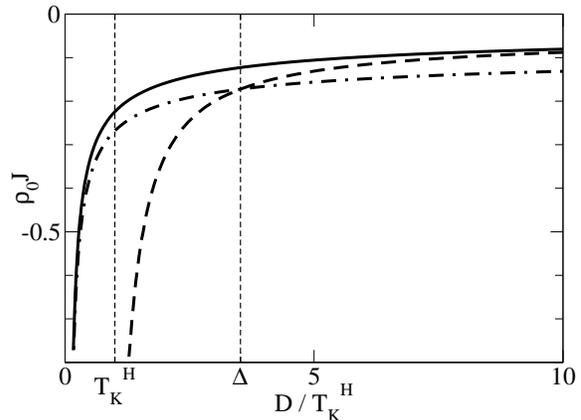}
  \caption{$D$ dependence of the scaled exchange coupling $J(D)$ of CS 
    Hamiltonian in the case of a ground state doublet and an excited quartet, 
    separated by the CF splitting $\Delta$.
    Solid curve is the 2nd-order Hanzawa scaled result. 
    Dashed and dot-dashed curves are the high-energy and low-energy 
    asymptotic scaling trajectories appropriate to a sixfold degenerate 
    and a doubly degenerate level, respectively.}
                         \label{fig:J2JHJ6}
\end{figure}

The scaling approach to the single-impurity models is explained 
in great detail in many papers
\cite{anderson.70,nozieres.80,yamada.84,hanzawa.85,hewson.93} 
and here we just use these results to calculate the thermopower.  
The scaling equations are generated  by reducing the 
conduction electron cutoff from $D_0$ to $D$ and simultaneously rescaling 
the coupling constant, $J(D)$, so as to keep the low-energy excitations 
of the total system unchanged.\cite{anderson.70}
The solution for the CSM, valid up to the 2nd order in renormalized 
couplings, reads\cite{yamada.84}   
\begin{equation}
                                    \label{hanzawa_2}
%\rho J
\exp \left( -\frac{1}{\rho_0 J} \right)
=
\left( \frac{T_K}{D} \right)^{m}
\left( \frac{T_K+\Delta }{D+\Delta } \right)^{M},
\end{equation}
where $T_K$ is the Kondo temperature, defined by $J_0$ and $D_0$. 
For a given set of initial parameters the renormalized coupling constant 
$J(D)$ is completely characterized by $T_K$. 
The result obtained for $m=2$, $M=4$, and $\Delta=330$ K 
is shown in Fig.~\ref{fig:J2JHJ6} 
as a function of $D/T_K^H$, where $T_K^H$ is defined below 
(see Eq.~\ref{T_K^H-T_K}).
For a given value of $D_0$ and $\Delta$, the Kondo temperature grows 
exponentially with $J_0$; for a given $D_0$ and $J_0$ the Kondo temperature  
grows with the CF splitting as $(\Delta/D_0)^M$.
The scaling trajectory defined by Eq.(\ref{hanzawa_2}) diverges at $D=T_K$, 
such that $J(T)$ can be used to renormalize the weak-coupling theories 
for $D\gg T_K$ only. 
Note that the CF splitting lowers the degeneracy of the {\it f} level and 
reduces the Kondo scale from $T_K^H$ to $T_K$, which extends 
the validity of the weak-coupling theory to much lower temperatures. 
The scaling trajectory $J(T)$, defined by Eq.~\ref{hanzawa_2}) and 
shown in Fig.~\ref{fig:J2JHJ6}, 
has two asymptotic regimes.\cite{hanzawa.85}    
The high-temperature asymptote, $J_H(T)$, is obtained from 
Eq.~(\ref{hanzawa_2}) by neglecting $\Delta$ with respect to $D$, 
which leads to   
\begin{equation}
                                    \label{T_K^H}
%\rho J_H
\exp \left( -\frac{1}{\rho_0 J_H}\right)
=
\left(\frac{T^H_K}{D}\right)^{N}, 
\end{equation} 
where $T^H_K$ is the high-temperature Kondo scale given by 
\begin{equation}
                                    \label{T_K^H-T_K}
(T_K^H)^{N}=(T_K)^m(T_K+\Delta)^M,
\end{equation}
and $N=m+M$. 
The same trajectory would be obtained for an effective sextet without 
the CF splitting, assuming the initial condition $J_H(D_0)=J_0$.

The low-temperature asymptote $J_L(D)$ is obtained by neglecting 
in Eq.~(\ref{hanzawa_2}) $T_K$ and $D$ with respect to $\Delta$, 
i.e., by neglecting the effects of the excited CF levels. 
This identifies $J_L(D)$ as the scaling trajectory of an effective doublet 
with Kondo scale $T_K$ and the effective coupling  
\begin{equation}
                                  \label{T_K^L}
%\rho J_L
\exp \left( -\frac{1}{\rho_0 J_L}\right)
=
\left(\frac{T_{K}}{D}\right)^{m}.    
\end{equation}
Note that the value of $T_K$ in the above expression is defined by 
Eq.~(\ref{hanzawa_2}) with the initial condition  $J(D_0)=J_0$. 
The initial condition $J_L(D_0)=J_0$ and Eq.~(\ref{T_K^L}) 
would lead to a Kondo temperature $T_K^L$ that would be much lower 
than $T_K$, i.e., the Kondo scale that enters Eq.(\ref{T_K^L}) 
is exponentially enhanced with respect to $T_K^L$ because of CF splitting.

Thus, the scaling analysis shows that the {\it f} level behaves at high temperatures 
as an effective sextet with the scaling trajectory $J_H(D)$ and Kondo 
scale $T_K^H$, while at low temperatures it behaves as an effective 
doublet with the scaling trajectory $J_L(D)$ and Kondo scale $T_L=T_K$. 
The scaling trajectory $J(D)$ of the CSM with the CF splitting  
interpolates between these two asymptotic regimes. 
Here, we discussed the scaling solution obtained by the 2nd-order 
perturbation expansion but the same behavior is obtained in the 
next leading order.\cite{hanzawa.85} 

%%%%%%%%%%%%%%%%%%%%%%%%%%%%%%%%%%%%%%%%%%
\subsubsection*{Renormalized perturbation expansion
\label{perturbation}}
%%%%%%%%%%%%%%%%%%%%%%%%%%%%%%%%%%%%%%%%
The scaling theory can be used to obtain a simple description of the 
overall magnetic and transport properties of the CSM.  
It provides the thermopower in the following way. 
We reduce the conduction-electron half-bandwidth from $D_0$ to 
$D=A \, k_B T$, where A is a numerical constant of the order of unity, 
and find the effective couplings $J(T)$, $V_m(T)$, and $V_M(T)$. 
Next, we notice that the reduction of the bandwidth does not change 
the form of the Hamiltonian, and that the form of the response 
functions obtained by perturbation expansion in terms of 
the effective coupling constants remain invariant with respect 
to the scaling.
Thus, to obtain the poor man's scaling result for the thermopower
we introduce the temperature-dependent cutoff $D=A\, k_B T$, 
find $J(T)$ from  Eq.(\ref{hanzawa_2}), and substitute this  
renormalized expansion parameter in the expressions 
\ (\ref{eq:Sbc})\ -\ (\ref{couplings}). 

To apply the scaling result to a specific Ce or Yb system we 
have to determine the degeneracy and the splitting of the CF 
levels, estimate the initial coupling constant (or $T_K$), 
and choose the cutoff constant A. The thermopower obtained 
by the renormalized perturbation theory depends on the 
magnitudes of the Kondo temperature, the potential scattering, 
the CF splitting, and the density of states of the conduction 
electrons at the Fermi level. 
These parameters are restricted to a rather narrow range by the 
requirements that the model explain not just the thermopower data  
but that it also leads to a consistent description of other 
thermodynamic and transport data of a particular system. 
We estimate the CF splitting and the degeneracies of the CF states 
by analyzing the neutron scattering and the high-temperature magnetic 
anisotropy data. 
The Kondo temperature $T_K$, i.e., the initial coupling $J_0$, 
can be obtained from the magnetic susceptibility data, 
provided we know how to extract the single-ion contribution 
from the experimental results\cite{aviani.01}. The high-temperature 
Kondo scale $T_K^H$ is then found from Eq.~(\ref{T_K^H-T_K}).
The relationship between $T_K$ and $J(T)$ depends to some extent 
on the scaling procedure and the value of the cutoff constant A. 
Our analysis of the thermopower of the CSM is based on the 2nd-order 
scaling equation\cite{yamada.84} and we use, for simplicity, $A=1$. 
The 3rd-order scaling equation\cite{hanzawa.85} and/or some larger 
values of A do not change the TEP in any qualitative way. 
The potential scattering shows up in the electrical resistance 
and typical results for the high-temperature phase\cite{scaling_V} 
 indicate rather large values for $\Vs$. 

\begin{figure}[tb]
    \epsfxsize=3.0in \epsffile{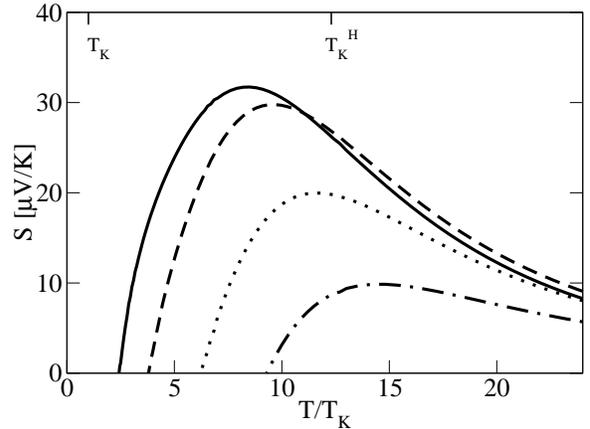}
    \caption{The thermopower is plotted as a function of reduced 
      temperature $T/T_K$ and for various values of the potential scattering. 
      Dot-dashed curve $\rho_0\Vs = -0.15$, dotted curve $\rho_0\Vs = -0.20$,
      dashed curve $\rho_0\Vs = -0.35$, solid curve $\rho_0\Vs = -0.5$.
      Here we used $T_K = 8$ K and $\Delta = 350$ K, 
      such that $T_K^H = 99$ K. 
      }
    \label{fig:TEP_V}
\end{figure}
\begin{figure}[tb]
    \epsfxsize=3.0in
    \epsffile{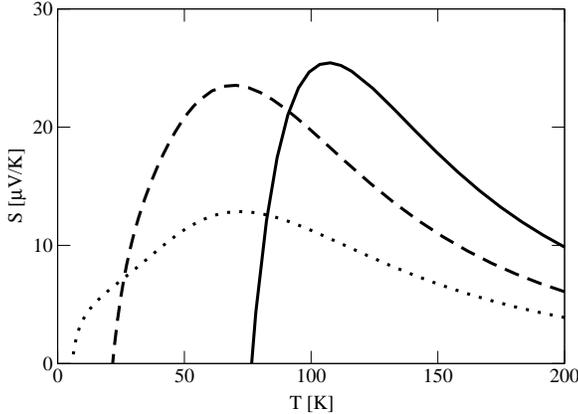}
    \caption{The thermopower is plotted as a function of temperature 
      for $\rho_0\Vs = -0.35$ and $\Delta = 350 \K$, 
      and for three values of $T_K$.
      Dotted curve $T_K = 2 \K$,
      $T_K^H = 63 \K$, dashed curve $T_K = 8 \K$, $T_K^H = 99 \K$,
      solid curve $T_K = 32 \K$, $T_K^H = 158 \K$.
      }
    \label{fig:SskaliranoTk}
\end{figure}
\begin{figure}[tb]
    \epsfxsize=3.0in 
    \epsffile{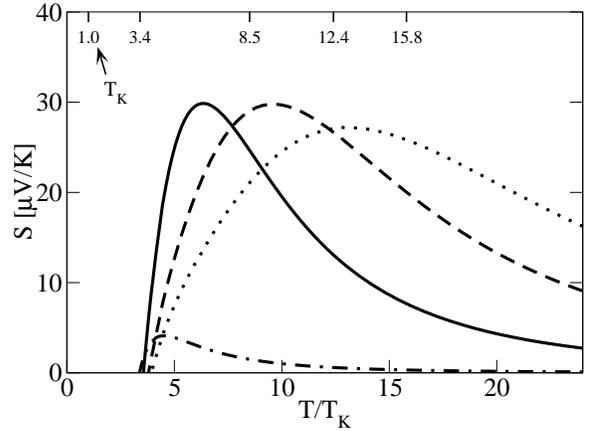}
    \caption{$S(T)$ is plotted as a function of reduced temperature 
      $T/T_K$ for $\rho_0\Vs = -0.35$  
      and for various values of the CF splitting.
      The curves are obtained for $T_K = 8 \K$, 
      and the dot-dashed, solid, dashed, and dotted curve 
      corrspond to 
      $\Delta =  50 \K$  ($T_K^H = 27 \K$),  
      $\Delta = 200 \K$ ($T_K^H = 68 \K$), 
      $\Delta = 350 \K$ ($T_K^H = 99 \K$), and 
      $\Delta = 500 \K$ ($T_K^H = 126 \K$), 
      respectively.}
    \label{fig:SskaliranoD}
\end{figure}

The thermoelectric power obtained for the doublet-quartet CF scheme, 
with $\Delta=350$ K and for several values of $T_K$ and  $\Vs$,
is shown in 
Figs.~\ref{fig:TEP_V}~-~\ref{fig:SskaliranoD}
The density of conduction states, which enters 
the thermopower calculations, is set to $\rho_0=2.2$ (eV)$^{-1}$.
Fig.~\ref{fig:TEP_V} shows the effect of $\Vs$ on $S(T)$ 
for $T_K=8$ K, $T_K^H=99$ K, and $\Delta=350$ K. 
The details of the shape of $S(T)$ depend in a rather complicated 
way on the relative magnitude of $T_K$, $\Vs$, and $\Delta$. 
For large $\Vs$ the value of $T_{max}$ is mainly 
determined by $T_K^H$, which is related to $T_K$ 
and $\Delta$ by Eq.~(\ref{T_K^H-T_K}). 
For small values of $\Vs$, the TEP changes sign before the 
full maximum can develop. The crossing temperature $T_x$, 
where $S(T)$ changes sign, i.e. $S(T_x)=0$, increases as $|{\Vs}|$ decreases. 

Fig.~\ref{fig:SskaliranoTk} shows $S(T)$ for $\Delta=350$ K, $\Vs=0.35$, 
and for 3 values of $T_K$. For such a large value of $\rho_0\Vs$, 
we find that the maximum of $S(T)$ is around $T_K^H$ and 
that $T_x$ scales approximately with $T_K$. 
Fig.~\ref{fig:SskaliranoD} shows the effect of $\Delta$ on $S(T)$ 
for $T_K=8$ K and $\Vs=-0.35$.
The position of $T_{max}$ is rather well described by $T_K^H$,
while the value of $T_x$ is not much changed by $\Delta$, 
as it mainly depends on $T_K$.
                                %                                %
The overall features of $S(T)$ shown in 
Figs.\ref{fig:SskaliranoTk} - \ref{fig:SskaliranoD} follow 
straightforwardly from  expressions (\ref{eq:Sbc}) - (\ref{eq:R_Delta}).
Eq.(\ref{eq:S_Delta}) contains the products of logarithmic functions 
and at high temperatures, where $J(T)$ changes very slowly, 
$S(T)$ reduces to the BC expression\cite{coqblin.76}. 
Eq.~(\ref{eq:S_Delta}) shows that the sign of the thermopower 
is determined by the temperature dependent factor ~$J(T)-\Vs$, 
and that $S(T)$ approaches zero for  $J(T)\simeq\Vs$.

However, to make a quantitative analysis of $S(T)$ close to $T_x$, 
we should not neglect the higher-order terms in the expansion
of transport integrals.
The 3rd-order perturbation theory provides an indication 
that $S(T)$ changes sign below $T_{max}$ and exhibits  
non-monotonic features. Since $S(T)$ must vanish at $T=0$,  
the finite value of $T_x$ must lead to a low-temperature minimum, 
which is here obtained from the exchange scattering, 
without invoking the interaction effects or long-range 
antiferromagnetic fluctuations.\cite{fisher.89} 

The properties of the model for $T\leq T_K$ can not be obtained 
from scaling. On the one hand, Eq.~(\ref{hanzawa_2}) 
is derived by assuming that thermal 
fluctuations do not excite too many conduction electrons or holes 
up to states near the (effective) band edges; 
this assumption does not hold for $T \leq T_K$, 
i.e., the scaling breaks down when the effective (renormalized) band 
edge is of the order $T_K$. 
On the other hand, the perturbation theory for the response functions 
breaks down close to $T_K$, because the effective coupling constants 
become too large for the lowest-order perturbation expansion to be valid. 
However, Fig.~\ref{fig:J2JHJ6} indicates that the breakdown of scaling 
occurs at temperatures such that  $J(T)\simeq J_L(T)$, and that the 
low-energy dynamics of the CSM is rather well described by an 
effective doublet or quartet. Thus, we neglect the CF splitting 
at low temperatures and calculate the thermopower of Ce and Yb intermetallics 
by using the methods that are more accurate than scaling. 

%%%%%%%%%%%%%%%%%%%%%%%%%%%%%%%%%%%%%%%%%
\subsection*{Low-temperature approximations
\label{low-T-TEP}}
%%%%%%%%%%%%%%%%%%%%%%%%%%%%%%%%%%%%%%%%
To discuss the low-temperature properties we start from the scaling 
result that the $4f^1$ or $4f^{13}$ states of the rare earth ions 
behave for $T\simeq T_K\ll T_K^H$ as effective levels with 
the degeneracy of the CF ground state. We assume that the coupling 
of such effective states to the conduction electrons characterizes 
the Kondo temperature which is the same as $T_K$ of the full CS model, 
and describe the low-temperature data by models without the CF splitting.
Since the reduction of temperature could lead to coherent 
scattering of conduction electrons on rare earth ions, stoichiometric 
compounds are treated differently from dilute alloys. 

%%%%%%%%%%%%%%%%%%%%%%%%%%%%%%%%%%%%%%%%
\subsubsection*{Periodic Anderson model}
%%%%%%%%%%%%%%%%%%%%%%%%%%%%%%%%%%%%%%%%

In ordered compounds, we assume that the coherence temperature 
of the lattice of rare earth ions, $T_0$, is not much different 
from $T_K$, the lowest Kondo temperature of the CS model.
Since the crossover from the full CF multiplet to an effective 
low-temperature CF state takes place much above $T_K$, 
we neglect the CF splitting in the coherent regime. 
Thus, for $T\leq T_0$ we describe the stoichiometric compounds 
by the periodic Anderson model (PAM) which has the degeneracy of the 
lowest CF state and adjust the effective parameters in such a way that 
the low-energy scale of the PAM coincides with $T_K$ of the CSM. 
Here we use the twofold degenerate PAM defined by the Hamiltonian  
\begin{eqnarray}
                                                    \label{PAM}
  H &&=    \sum_{i\sigma} E_f a^{\dag}_{i\sigma} a_{i\sigma}
         + \sum_{k} \sum_{\sigma} \epsilon_k\, c^{\dag}_{k\sigma} c_{k\sigma}
                                                         \\
    &+& (V \sum_{k} \sum_{i\sigma} e^{i k R_i} 
          c^{\dag}_{k\sigma} a_{i\sigma} + h.h.) 
     + U \sum_{i} a^{\dag}_{i\uparrow} a_{i\uparrow} 
                   a^{\dag}_{i\downarrow} a_{i\downarrow}, \nonumber 
\end{eqnarray}
where $\sigma$ labels two spin states of the localized level, 
$E_f$ is the unrenormalized position of the effective CF ground state, 
$V$ is the hybridization, and U is the Coulomb repulsion. 
We consider the case where the {\it f} level is below the Fermi level, 
and the width of the {\it f} band $W$ is smaller than the 
on-site correlation. 

A simple but non-trivial approximation to study the PAM is provided by the
2nd-order perturbation theory (SOPT) with respect to the correlation $U$.
The SOPT, which may be regarded as the simplest extension of Hartree-Fock
theory, properly reproduces Fermi liquid properties and  mass enhancement,
and it reproduces a characteristic low-temperature scale $T_0$. A
disadvantage of the SOPT is the fact that this $T_0$ has not the proper
(non-analytic) dependence on the model parameters (hybridization $V$ and
correlation $U$, in particular), as it should according to the
Schrieffer-Wolff transformation. But on the other hand, arbitrarily large mass
enhancements and any value of $T_0$ can be achieved in SOPT by choosing
appropriate values for the parameters $V$ and $U$. Therefore one can choose
the parameters of the effective PAM in such a way that $T_0$ corresponds
to the true $T_K$ of the CS model. The SOPT is most easy to apply in the
limit of high dimensions $d \rightarrow\infty$,\cite{MetzVoll} 
because then the self-energy is
{\bf k}-independent (site-diagonal), which is already a good additional
approximation for the realistic dimension $d=3$. Consistent with the site
diagonality of the self-energy is the vanishing of vertex corrections. Then
the transport coefficients can be calculated from the Kubo formula. The
static (zero-frequency) conductivity is given by\cite{czycholl.91}

\begin{equation}
\sigma_{xx} = \frac{e^2 a^{2-d}}{2\pi \hbar} t^2 \int dE
\left(-\frac{df}{dE}\right) L(E), 
\label{pamconductivity}
\end{equation}
where $a$ is the lattice constant, $d$ the dimension, $t$ the hopping
matrix element of the band electrons, $f(E)$ the Fermi function, and the
function $L(E)$ is given by
\begin{equation}
L(E) = \frac{2}{N} \sum_{\vec k \sigma} 
\left(\mbox{Im} G^c_{\vec k\sigma}(E+i0)\right)^2.
\end{equation}
The function 
\begin{equation}
G^c_{\vec k \sigma}(z) = \frac{1}{z - \frac{V^2}{z - E_f - \Sigma(z)} -
\varepsilon_k}
\end{equation}
is the band electron Green's function of the PAM, and 
$\Sigma(z)$ the {it f}-electron self-energy, which we determine 
approximately within SOPT. 
The TEP is also determined by the function $L(E)$ according to
\begin{equation}
S = \frac{\int dE \left(-\frac{df}{dE}\right) (E-\mu)L(E)}{eT
\int dE \left(-\frac{df}{dE}\right)L(E)}
\label{pamthermop}
\end{equation}

As mentioned above, the SOPT yields a characteristic low-temperature scale
$T_0$ which can be determined or defined as follows. The {it f}-electron
density of states (DOS) at the chemical potential $\mu$ is strongly 
temperature-dependent and decreases as temperature increases 
on the scale $T_0$ towards an asymptotic T-independent value.
Therefore, $T_0$ can be defined as the half-width of the T-dependent part of
the {it f}-DOS at $\mu$. When calculating the T-dependence of the resistivity
$R(T) = 1/\sigma_{xx}$ from (\ref{pamconductivity}), one obtains for most
choices of the parameters the following characteristic behavior: 
A residual resistivity approaching zero for $T \rightarrow 0$, 
a $T^2$ dependence for very low $T$ as expected for Fermi
liquids, a nearly linear increase with increasing $T$ for $T < T_0$, a
maximum of $R(T)$ exactly at $T_0$, and an $R(T)$ decreasing with increasing
$T$ (and thus behaving similarly as in the case of incoherent scattering
from magnetic impurities) for $T > T_0$. 

%\vspace{1cm}                   %
\begin{figure}[tb]
  \begin{center}
   \epsfxsize=3in \epsffile{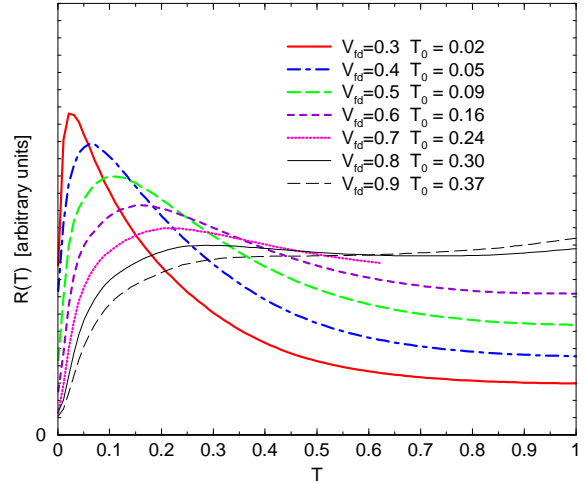}    
     \caption{The resistivity of the periodic Anderson model 
             is shown for various values of the hybridization 
             parameter as a function of temperature measured 
             in units of the band-width W. 
The total number of electrons per site is  $n_{tot} = 2.2$ 
and the number of {\it f} holes is $n_f^{hole} = 0.9$.
             The value of the Coulomb correlation is $U/W=1$ 
             and the corresponding low-energy scale $T_0$ 
             is indicated in the figure.}
    \label{pam_rho.eps}
  \end{center}
\end{figure}
As an example, we consider the model with more than one {\it f} electron 
per site. This would correspond to Yb systems in which the 
number of {\it f} holes is restricted by large Coulomb correlation 
to  $n_f^{hole}\leq 1$. The numerical results are shown in 
Fig.~\ref{pam_rho.eps}, where $R(T)$ is plotted versus temperature 
for the model with the total number of electrons per site $n_{tot}=2.2$,  
the occupancy of the {\it f} holes $n_f^{hole}=0.9$, 
and for different values of hybridization $V$.    
For fixed other parameters the hybridization $V$ determines the 
low-temperature scale $T_0$, which is indicated in the figure.
The calculations are performed for $U/W=1$.
Remarkably, as long as there is a maximum in $R(T)$, it is very 
close to $T_0$ as determined by the {\it f} DOS criterion described above. 
For too large $V$ and $T_0$ (and the corresponding less strong mass 
enhancement) there is no longer a true maximum in $R(T)$ 
but only a plateau behavior. 

%\vspace{1cm}                   %
\begin{figure}[tb]
  \begin{center}
   \epsfxsize=3in \epsffile{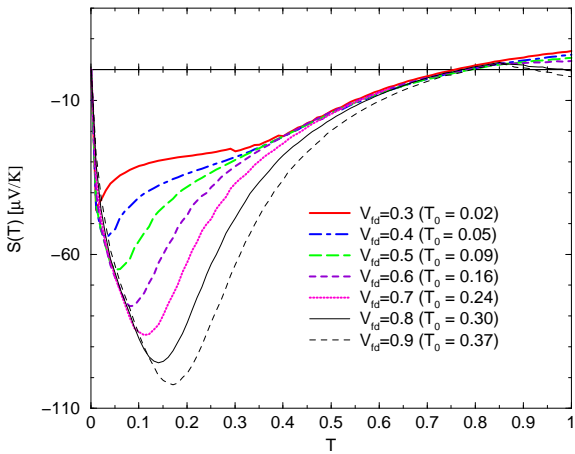}
    \caption{The thermopower of the periodic Anderson model 
             is shown for various values of the hybridization 
             parameter as a function of temperature, measured 
             in units of the band-width W. 
             The total number of electrons per site is $n_{tot}=2.2$ and 
             the number of {\it f} holes is $n_f^{hole}=0.9$.
             The value of the Coulomb correlation is $U/W=1$
             and the corresponding low-energy scale $T_0$ is indicated 
             in the figure.}
    \label{pam_tep.eps}
  \end{center}
\end{figure}
Corresponding results for the thermopower are shown in Fig.~\ref{pam_tep.eps}.
Note that one obtains $S(T)$ in its natural units 
(using $k_B/e\approx 86 \mu V/K$) from (16), as the ''arbitrary'' 
units (prefactor of (13)) cancels due to the quotient. 
%%%%%%%%%%% %where $S(T)$ is measured in units of k$_B$/e $\approx 86 \muV/K. 
The initial sign of $S(T)$ depends on the slope of $L(E)$ at the Fermi level  
and the temperature variation of $S(T)$ reflects the structure 
of the DOS within the Fermi window. 
Fig.~\ref{pam_tep.eps} shows typical results for $S(T)$ 
of Yb systems, with a large negative low-temperature peak.
Obviously $S(T)$ is strongly temperature-dependent and its absolute 
value is very large, namely of the magnitude 50 - 100 $\mu$V/K. It has an
extremum (negative minimum in the plot) at a temperature $T_1$, which scales
linearly with $T_0$, i.e., $T_1 = a T_0$. 

For the parameters used for the figure we have $a \approx 0.5$, but the 
exact value of $a$ depends on other parameters (U, n$_f$ etc.) as well. 
This extremum in $S(T)$ also exists in
the situation when the resistivity $R(T)$ exhibits no maximum but only a
plateau behavior. The absolute value $|S(T_1)|$ at the extremum can even 
increase with increasing $T_0$, i.e., according to this result 
it is not necessarily the most ''heavy'' fermion systems that exhibit the
largest values of the TEP. In any case, in the low-temperature regime, 
$T\leq T_0$, this approach yields a TEP of the correct absolute magnitude and
qualitative behavior, in particular an extremum, which is characteristic
for the  TEP  experimentally observed in many heavy fermion systems. Of
course for intermediate and high $T$ the features due to the CF splitting
cannot be reproduced within this SOPT treatment of the PAM, as only the
twofold degenerate PAM was used, which has no higher CF-split {\it f} levels
included.      

%%%%%%%%%%%%%%%%%%%%%%%%%%%%%%%%%%%%%%
\subsubsection*{Single-impurity case}
%%%%%%%%%%%%%%%%%%%%%%%%%%%%%%%%%%%%%%
The transport properties of dilute Ce and Yb alloys with 
the doubly degenerate CF ground state are obtained at temperatures 
below $T_K$ from the single-impurity spin-1/2 Anderson model. 
We assume that the number of {\it f} electrons (holes) is slightly 
above (below) one for each Ce (Yb) impurity, and we consider 
an asymmetric model. The effective parameters of the SIAM are adjusted 
in such a way that the width of its Kondo resonance coincides with the 
Kondo temperature $T_K$ of the CSM with CF splitting.

The TEP of a spin-degenerate SIAM, in the absence of the 
non-resonant scattering channels, has been calculated by 
various methods and is well understood. We are interested 
in the low-temperature behavior obtained for $U/\pi\Gamma\gg 1$, 
where U denotes the f-f correlation and $\Gamma=\pi\rho_F\V^2$ is the 
half-width of the {\it f} level due to the hybridization with conduction states.  
(Here, $\rho_F$ is the density of conduction states at the Fermi level.) 
The numerical renormalization group calculations(NRG)  
of Costi, Hewson and Zlati\'c\cite{zlatic.93,costi.94a}  
(in what follows referred to as CHZ) 
and the perturbation theory\cite{zlatic.82} 
show that $S(T)$ is closely related to the Kondo resonance. 
In the absence of any potential scattering the behavior of the TEP 
follows simply the temperature dependence of the {it f}-electron 
spectral function within the Fermi window. 
The results for the 4f$^1$ electron (and analogous results 
for the 4f$^{13}$ hole) can be summarized as follows. 
\cite{zlatic.93,costi.94a,zlatic.82} 
Close to  $T=0$ the system is in the Fermi liquid (FL) regime,
which is characterized by an asymmetric Kondo resonance of the 
width $T_K$, centered above the Fermi energy, $E_F$.
This leads to a typical Fermi liquid power law,  
\begin{equation} 
S(T)
=
\frac{\pi^2 k_B}{3|e|}
\frac{T}{T_K} 
cot \;\; \eta_0(E_F) ,
\label{resonant_SIAM}
\end{equation} 
where $\eta_0(E_F)$ is the resonant phase shift due to the 
scattering of conduction electrons on the {\it f} state. 
It is related to $n_f$ by the Friedel sum rule, 
$\eta_0(E_F)=\pi n_f/2$, so that for $n_f<1$ ($n_f>1$) 
the initial slope of the TEP is positive (negative). 
In heavy fermions, where temperatures of the order of $T_K$ are 
easily accessible, the TEP grows rapidly with temperature 
($\pi^2k_B/3|e|=284\,\mu V/K$) and can assume giant values. 
At elevated temperatures the Fermi liquid behavior breaks down, because 
the spectral weight is transferred out of the low-energy region and 
the Kondo resonance disappears. 
For $T>T_K$, the system enters the loca- moment (LM) regime. Here,  
for $n_f>1$, the maximum of the spectral function shifts from above 
to below $E_F$ and the TEP is negative. For $n_f<1$, we find that 
in the LM regime the TEP is positive. 

These simple features are modified in the presence of non-resonant 
scattering at the impurity site, because the interference between the 
resonant and the non-resonant channels leads to vertex corrections, 
which have drastic effects on the thermopower. 
These effects have been analyzed in detail in CHZ, and here we use that  
theory to study the changes in $S(T)$ induced by a small variation 
in $n_f$. Such changes in $n_f$ could be induced by pressure or 
doping (chemical pressure). The TEP is still given by the expression   
(\ref{pamthermop}), but the evaluation of the Kubo formula 
gives for $L(\omega)$ the result\cite{costi.94a} 
\begin{equation}
  \frac 1 {L(\omega)} = \frac 1 {L_0(\omega, T)} \left[
        \cos 2\eta_1 - \frac{\mbox{Re}\;G(\omega)}
                            {\mbox{Im}\; G(\omega)} \sin 2\eta_1
        \right]
        + \rho_n.
\end{equation}
Here,  $\eta_1$ is the non-resonant phase shift, 
$\rho_n\simeq sin^2(\eta_1)$ is the residual resistance 
due to the non-resonant scattering,  and 
$L_0(\omega)$ is the transport relaxation time due to the 
resonant scattering,  
\begin{equation}
                 \label{L_0}
  \frac 1 {L_0(\omega)} = - \frac {\Gamma} \pi \mbox{Im}\; G(\omega), 
\end{equation}
(for details of derivation see CHZ).
The resonant Green's function in Eq.~(\ref{L_0}) has been calculated by 
the NRG methods\cite{costi.94a} and the results were used to study 
the thermopower at the crossover from above to below $T_K$. 
However, in the presence of CF splitting we do not need 
the resonant Green's function of the spin-1/2 problem for 
temperatures much above $T_K$. In the Fermi liquid regime it 
is sufficient to evaluate  $G(\omega)$ by an approximate 
method\cite{martin-rodero.86} that interpolates between the 2nd-order 
weak-coupling expression (in what follows referred to as MR) 
and the exact atomic limit.
This simple approximation is quite accurate for $T\leq T_K$;
it gives the spectral function and the thermoelectric power similar 
to the NRG results of CHZ even for $U/\pi\Gamma\gg 1$. 
The approximation becomes unphysical for $T\gg T_K$ but here we 
are not interested in such high temperatures where the spin-1/2 
model becomes inadequate anyway. 
The approximation is based on the Dyson equation, 
\begin{equation}
                 \label{G_MR}
  G(\omega) = \frac 1 
  {\omega - \epsilon_l + i \Gamma - U n_f - \Sigma(\omega)}, 
\end{equation}
with $\epsilon_l$ determined iteratively in such a way that the number of
electrons per spin, 
\begin{equation}
  \tilde n = \int d\omega f(\omega) 
  \left( - \frac 1 \pi \mbox{Im} \;G \right), 
\end{equation}
equals $n_f$. 
The self-energy $\Sigma(\omega)$ is 
defined by the Martin-Rodero interpolation, 
\begin{equation}
  \Sigma(\omega) = \frac{\tilde \Sigma^{(2)}(\omega)}
  {1 - \frac{(1 - n_\sigma)U + \epsilon_l - \epsilon_0}
            {n_\sigma (1 - n_\sigma) U^2} 
    \tilde \Sigma^{(2)}(\omega))}, 
\end{equation}
where, $\epsilon_0$ is the energy of the virtual bound state 
of the $U=0$ Anderson model with $n_f$ electrons, 
and $\tilde \Sigma^{(2)}(z)$ is the 2nd-order self-energy
calculated with the unperturbed propagators of this auxilliary 
$U=0$ model (for details see Ref. MR). 

\begin{figure}[tb]
  \begin{center}  
   \epsfxsize=3in \epsffile{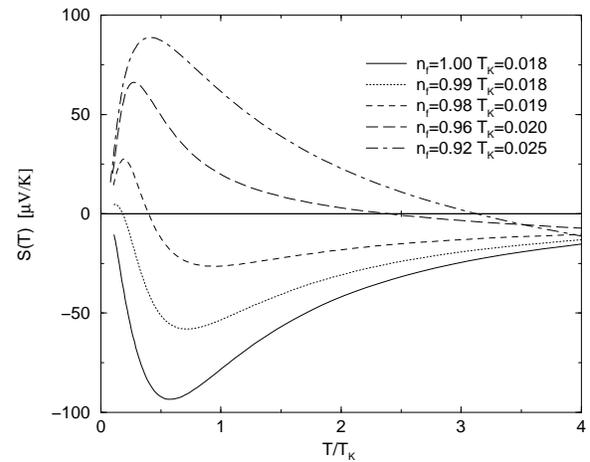}
    \caption{The thermopower of the single-impurity Anderson model 
             is shown for various {\it f} occupation 
             as a function of temperature, measured in units of $T_K$. 
             The value of the Coulomb correlation is $U/\pi\Gamma=8$
             and the corresponding Kondo scale $T_K$ is indicated 
             in the figure.}
    \label{siam_U=8_tep.eps}
  \end{center}
\end{figure}

The thermopower obtained by this approximation is plotted 
in Fig.~\ \ref{siam_U=8_tep.eps} as a function of temperature and for 
several values of $n_f$. We consider the case of strong correlation,  
$U/\pi\Gamma=8$, and assume a small non-resonant scattering, $\eta_1=-0.1$ 
(the effect of $\rho_n$ on $S(T)$ is neglected). 
The value of $T_K$ (in units of $\pi \Gamma$) is estimated from 
the {\it f} spectral function that we obtain from the Dyson equation (\ref{G_MR}).
We use two different estimates: 
(i) we associate $T_K$ with the full width at half the maximum (FWHM) 
    of the Kondo resonance at $T=0$, and 
(ii) we identify $T_K$ as the half-width of the temperature-dependent 
     part of the {it f}-electron spectral function at the chemical potential. 
The value of $T_K$ obtained by either method is almost the same, 
and we find that the reduction of $n_f$ enhances $T_K$. 

%\vspace{1cm}                   %
The $n_f=1$ curve corresponds to an extreme Kondo limit, where $S(T)$ 
goes rapidly to a large negative peak at about $T\simeq T_K/2$, 
and then decreases gradually but remains negative within the physically 
relevant temperature range $T\leq T_K$. 
The effect of the non-resonant phase shifts is very important, because 
at half-filling ($n_f=1$) the model is  electron-hole symmetric and in 
the absence of the non-resonant phase shifts the thermopower vanishes. 
The  $n_f=0.98$ and $n_f=0.99$ curves are still Kondo-like, but the 
reduction of $\eta_0$ from $\pi/2$ gives rise to a positive initial 
slope of $S(T)$, such that $S(T)$ develops a well defined positive 
maximum before decreasing sharply to negative values 
for temperatures between $T_K/2$ and $T_K$. 
Finally, the  $n_f=0.96$ and $n_f=0.92$ curves remain positive  
for $T\leq T_K$. Further reduction of $n_f$ does not bring any 
new qualitative features, except that the maximum of $S(T)$ becomes 
higher and shifts to higher temperatures. 
By changing the other parameters, like the correlation or the 
non-resonant phase shifts, we find a similar set of curves, that  
could be classified in the same way as the curves shown in 
Fig.~(\ref{siam_U=8_tep.eps}).

The low-temperature thermopower results shown in Fig.~(\ref{siam_U=8_tep.eps}) 
exhibit all the features found in dilute Ce alloys for $T\leq T_K$. 
Reflecting the curves on the $S=0$ axis, gives $S(T)$ that 
correspond to the Yb alloys, with one magnetic {\it f} hole. 

%%%%%%%%%%%%%%%%%%%%%%%%%%%%%%%%%%%%%%
\subsection*{Discussion of different cases (a) to (e) and 
conclusion\label{discussion}} 
%%%%%%%%%%%%%%%%%%%%%%%%%%%%%%%%%%%%%%

The overall temperature dependence of the TEP of Ce and Yb 
intermetallics is found by interpolating between the results 
for the Anderson model, valid below $T_K$, and the CSM results,  
valid above $T_K$.  
The essential features of $S(T)$ are due to the Kondo effect modified  
by CF splitting, and various shapes are the consequence of different 
energy scales that characterize the low- and high-temperature behavior. 
The details, however, depend not just on the single parameter $T_K$, 
but on other parameters of the model as well. Here we discuss, for simplicity, 
the thermopower due to the cerium ions with a single electron 
in the 4f$^{1}$ configuration.   
The results pertaining to ytterbium ions with a single hole in 
the 4f$^{13}$ configuration are obtained by reflecting the $S(T)$ 
on the temperature axis. 

%%%%%%%%%%  a-b %%%%%%%%
In (a)- and (b)-type systems a large CF splitting of Ce {\it f} states generates 
two distinct energy scales, which are seen as two Kondo temperatures 
of the CS model, $T_K$ and $T_K^H\gg T_K$.   
The high-temperature Kondo scale describes a sixfold degenerate 
CF multiplet with a large thermopower. Typically, $T_K^H$ 
is between 30 and 100 K and the maximum value of the thermopower 
can exceed 150 $\mu$V/K. 
The low-temperature scale, defined by $T_K$, is of the order 
of 1 to 10 K. A twofold or fourfold degenerate CF ground state  
gives rise, below $T_K$, to an additional structure in $S(T)$.
In periodic systems, the onset of coherence reverses the sign of the 
thermopower below $T_K$ and gives rise to a large negative peak 
around $T_0/2\simeq T_K/2$. The behavior of $S(T)$ in the $T=0$ limit 
depends on the band filling and the value of the f-d hybridization: 
for systems close to the electron-hole symmetry $S(T)$ 
is positive for $n_f < 1$ and negative for $n_f > 1$.
Combining these results with the $T\leq T_K$ results one obtains  
the thermopower of the (a) or (b) type.
In dilute Ce alloys the sign of the low-temperature thermopower 
is determined by the non-resonant phase shifts.  
Type (b) behavior is obtained here for a ground CF-level that gives rise 
to a positive thermopower around $T_K/2$ and a negative one above $T_K$.  
For sufficiently different values of $T_K$ and $T_K^H$ the $S(T)$ 
has two well resolved positive peaks separated by a negative minimum. 
Changing the sign of the non-resonant phase shifts one reverses 
the sign of the low-temperature thermopower and finds the (a)-type 
thermopower: there is a deep negative minimum around $T_K/2$, followed  
by a positive maximum around $T_K^H$. 
The type-(a) and type-(b) systems can be classified as Kondo systems 
with large CF splitting. The crossover from the high- to the low-temperature 
fixed point is accompanied by the sign change of the TEP. 

%%%%%%%%%%  c %%%%%%%%
The type-(c) behavior can arise in two ways. In the first case, 
the low- and high-temperature maxima overlap, giving a two-hump structure. 
That is, the CF splitting is not large enough to generate sufficiently 
different Kondo scales for the negative minimum to occur. 
In the second case, the CF splitting might be large but the 
potential scattering is also large enough to prevent the thermopower 
due to the CF ground state from changing sign for $T\leq T_K$. Thus, 
the crossover from the low- to high-temperature fixed point occurs 
at  $T_K$ without the change of sign of $S(T)$. 
The two LM regimes are still well resolved but $S(T)$ is always positive.  

%%%%%%%%%%  d %%%%%%%%
The type-(d) thermopower is obtained for Kondo systems with small CF 
splitting, such that the low- and high-temperature thermopower peaks overlap,  
and we see but a single peak with, perhaps, a shoulder on the low-temperature 
side. In the absence of the CF splitting there is just a single 
thermopower peak at about $T_K/2$. The sixfold degenerate 
4f$^1$ electron of Ce ions has a large Kondo scale  
and huge positive thermopower. (The eightfold degenerate 
4f$^{13}$ multiplet of Yb ions could lead to even larger negative 
thermopower.) 

%%%%%%%%%%  e  %%%%%%%%
In valence fluctuators, the {\it f} level is close to the chemical potential  
and Ce ions fluctuate between two configurations.  
The thermopower of such a system increases monotonically to a 
large positive (or negative) value. The characteristic energy scale is 
defined by the width of the {\it f} level, and  $S(T)$ might reach 
very large values.

Finally, we remark that in Kondo systems with Ce ions the application 
of pressure (or positive chemical pressure) enhances $T_K$ and 
transforms an (a)- or (b)-type thermopower to a (c)- or (d)-type one. 
The negative (chemical) pressure reduces $T_K$ and gives rise to 
a reversed behavior. 
In Kondo systems with Yb ions the pressure and chemical pressure 
have the opposite effect than in Ce systems. We hope this simple 
explanation of the overall behavior of the thermopower of Ce and Yb 
systems to facilitate the search for optimal thermoelectrics. 

Acknowledgment. 
We acknowledge useful comments from I. Aviani, B. Horvati\'c, A. Hewson, 
J. Freericks, M. O\v cko, C. Geibel J. Sarrao, and F. Steglich.
One of us (V.Z.) gratefully acknowledges the financial support 
from the Swiss NSF (project number 7KRPJ065554-01/1). 
%%%%%%%%%%%%%%%%%%%%%%%%%%%%%%%%%%%%%%

%\newpage

%\newpage  


\begin{thebibliography}{99}
\bibitem{scaling_V}
The total potential scattering, $\Vs=V_0-J_0<n>$, is scaling invariant, 
because the scaling equations involve only the traceless part of 
the Hamiltonian\cite{nozieres.80}.
\end{thebibliography}
\end{document}